\documentclass[12pt]{article}
\usepackage{amsmath,amssymb}
\usepackage{graphicx}

\def\be{\begin{equation}}
\def\ee{\end{equation}}
\def\bea{\begin{eqnarray}}
\def\eea{\end{eqnarray}}
\def\ben{\begin{eqnarray*}}
\def\een{\end{eqnarray*}}

\def\nn{\nonumber\\}
\def\tr{\rm tr}
\def\delete#1{}

\begin{document}

\title{A Classical Solution of Massive Yang-Mills Fields}

\author{Tsuguo MOGAMI
	\thanks{e-mail: mogami290@gmail.com}\\
	Niikura Institute for Mathematical Physics,\\
	Shimoniikura 2-4-7, Wako-shi,
	Saitama, 351-0111 Japan \\
}
\date{June 1, 2014}			

\maketitle


\begin{abstract}
Recent researches on the solution of Schwinger-Dyson equations, as well as lattice simulations of pure QCD, suggest that the gluon propagator is massive.
In this letter, we assume that the classical counterpart of this massive gluon field may be represented with the equation of motion for Yang-Mills theory with a mass term added.  A new classical solution is given for this equation.
It is discussed that this solution may have some role in confinement.
\end{abstract}

These days, evidences are accumulating that the lattice QCD, which is equivalent to the real QCD, dynamically acquires mass from analytical and lattice studies\cite{BIMS, Nat, Agu} (and see \cite{Cor09} for a list of references).
The analytical studies with the Schwinger-Dyson equation (SDE) nicely agree \cite{Nat, Agu} with the lattice data.
Those SDE analyses are based on Landau gauge.

In this letter, the classical counterpart of massive gluons in the Landau gauge is considered.  In the next section, a new classical solution is given for the equation of motion, and its relation to confinement is discussed.
The mechanism here will equally be valid as far as the theory has a mass but the color symmetry is unbroken.  

Though this theory is intended to give insight into real QCD, a non-QCD toy model is additionally analyzed in section 2.
This model is not for the real world but for facilitating analysis in a toy world.  This model shows interesting behavior of the mass gap and absence of colored states.

\section{ A classical Solution }

\begin{figure}[tbp]
	\centering
	\includegraphics[width=90mm]{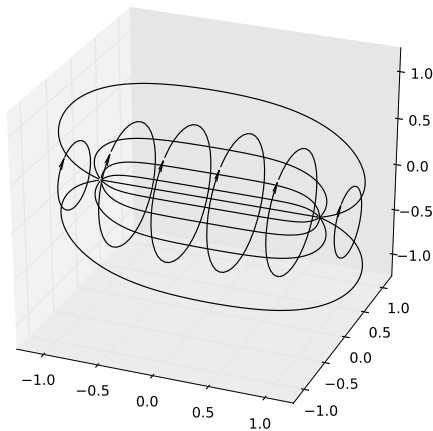}
	\caption{\small The electric and magnetic field between the quarks with their separation $2a = 2$.  The mass $m = 4$.  Electric flux on the vertical plane and magnetic flux on a tube for the zero-th order classical solution are shown.}
	\label{fig:em4}
\end{figure}

We assume that the classical limit of the massive gluons is represented by the equation of motion with a mass:\be
		- D_{\nu } F_{\mu \nu } - m^{2} A_{\mu } = 0 ,
		\label{eqm}
\ee
where \be
	F_{\mu \nu }^{a} = \partial _{\mu } A_{\nu }^{a} - \partial _{\mu } A_{\nu }^{a} + g f_{abc} A_{\mu }^{b} A_{\nu }^{c} ,
\ee
and $D_{\mu }$ is the covariant derivative.  The Landau-gauge condition $\partial _{\mu } A_{\mu }^{a} = 0$ follows from this equation by applying $\partial _{\mu }$ on the equation.
At the zeroth order in $g$, we find the following solution\delete{, which has a string between the quarks,} \bea
	&& A_{0 }= 0,		\\
	&& \vec A = \omega ^{-1}(\lambda _{2} \cos \omega t +\lambda _{3} \sin \omega t) \vec H_{0}(z, r) ,
	\label{H0}
\eea
where \be
	\vec H_{0} = \Big( \partial _{x},\ \partial _{y},\ \partial _{z}  - (m^{2} -\omega ^{2}) \int dz \Big) \phi (z, r) ,
\ee
with $r_{1} = \sqrt {(z - a)^{2} + r^{2}}, r_{2} = \sqrt {(z + a)^{2} + r^{2}} $ and $\phi  = q e^{- \sqrt {m^{2} -\omega ^{2}} r_{1} }/r_{1 }
- q e^{- \sqrt {m^{2} -\omega ^{2}} r_{2}} /r_{2} $.
The electric and magnetic field is \bea
	\vec E^{a} &=& -\vec \partial A_{0}^{a} - i g f_{abc}\vec A^{b} A_{0}^{c} - \dot{\vec A}^{a} ,		\\
	\vec B^{a} &=& \vec \nabla \times \vec A^{a} .
\eea
This solution may be interpreted as having two point charges placed at $ z = a$ and $z = -a$ with the opposite signs.  This electric field has no divergence ${\rm div} \vec E = {\rm div}\vec H = 0 $ except for the points on the charges with the magnitude $q$.  The electric flux looks like that of magnetic field in superconducting material.  The color of the magnetic field and the electric field are 90 degrees different in the space of color, and the magnetic flux keeps inducing the electric flux.  The shape of electric flux and magnetic field are shown in figure \ref{fig:em4}.
The electric flux looks like a tube elongated in $z$-direction.  When we pull the electric charges apart, the tube will get longer but it won't get thicker and be about $m^{-1}$.  Therefore, a constant tension will occur between the charges, which is quite consistent with stringy picture\cite{string} of confinement.

The zeroth-order solution above may be improved to all the orders perturbatively.  The equation of motion (\ref{eqm}) in a expanded form is \be
	\Box A^{\mu a} - \partial _{\mu } \partial _{\nu }A^{\nu a} - m^{2} A^{\mu a} = g f_{abc} \partial ^{\nu } A_{\mu }^{b} A_{\nu }^{c} + g f_{abc} A_{\nu }^{b} F_{\mu \nu }^{c} .
	\label{eqm2}
\ee
We can perturbatively obtain the solution by repeatedly applying this equation.
In addition, we have $ \partial _{\mu }A^{\mu } = 0 $ not only for the zeroth-order solution (\ref{H0}) but also to all the orders in $g$.  The right hand side of eq.(\ref{eqm2}) gives zero if we apply $\partial _{\mu } $ on it, since the first term gives $ \partial ^{\mu } f_{abc} \partial ^{\nu } A_{\mu }^{b} A_{\nu }^{c} = 0 $, and the second term $ f_{abc} A_{\nu }^{b} F_{\mu \nu }^{c} \equiv  J_{\mu ^{a} }$ will become \be
	\partial ^{\mu } J_{\mu ^{a} } 
	= (\partial ^{\mu } A^{\nu a})(\partial _{\mu } A_{\nu }^{a} - \partial _{\nu } A_{\mu }^{a}) + A^{\nu a}\partial ^{\mu }(\partial _{\mu } A_{\nu }^{a} - \partial _{\nu } A_{\mu }^{a}) + g \partial ^{\mu } A^{\nu a} f_{abc} A_{\mu }^{b} A_{\nu }^{c} .
\ee
Using symmetry, the equation of motion and Jacobi's identity, we have \bea
	 \partial ^{\mu } J_{\mu }^{a} &=& g A^{\nu a} f_{abc}(\partial _{\mu } A_{\nu }^{b}A_{\mu }^{c} + A_{\mu }^{b} (\partial _{\mu } A_{\nu }^{c} - \partial _{\nu } A_{\mu }^{c}) + g A^{\mu b} f_{cde} A_{\nu }^{d} A_{\mu }^{e} + m^{2}A^{\nu a})	\nn
		&& + g \partial ^{\mu } A^{\nu a} f_{abc} A_{\mu }^{b} A_{\nu }^{c} 
	= 0 .
\eea
Then $\partial _{\mu }A^{\mu } = 0$ follows at the $(n+1)$-th order, if $\partial _{\mu }A^{\mu } = 0$ holds at the $n$-th order in $g$.

This classical solution gives a picture that $E$ and $B$ are rotating within a color plane that includes $\lambda _{2}$ and $\lambda _{3}$ direction, and quark charges are rotating too.  Quantum mechanically, a quark must change its color after emitting a gluon.  This should have been the reason why it was difficult to understand the static force between confined quarks in the analogy of electric force.  Further this picture clarifies why non-Abelian nature is essential for confinement.

In the solution of massive Yang-Mills theory such as Weak theory, the solution does not form conserved flux tubes and its electric field vanishes at longer distance.  The reason why we don't have such a decaying solution is because we don't have current conservation $ \partial ^{\mu } J_{\mu }^{a} = 0 $ for the theory with broken global symmetry.

This confinement picture will be valid for all the theories that have the same equation of motion classically, including real QCD and lattice QCD.

\section{Explicit mass introduction}
Here we additionally present a toy but quantum model, in which the mechanism presented in the previous section holds and easier to analyze.  We consider a theory with its Lagrangian: 
\be
		{\cal L} \equiv  \frac{1}{4} (F_{\mu \nu }^{a})^{2} + \frac{1}{4} \tr |\partial _{\mu }\phi - ig A_{\mu }^{a} \tau ^{a} \phi |^{2} - \frac{\lambda }{4} \tr (\phi ^{\dagger }\phi )^{2} + \frac{\lambda M^{2}}{2} \tr \phi ^{\dagger }\phi ,
\ee
and with a spontaneous symmetry break of \be
		\langle \phi _{ij}\rangle  = v \delta _{ij},
		\label{VEV}
\ee
where $\phi $ is a complex valued $N \times N$ matrix field, and its left index couples to the gauge field but its right index does not.  The gauge symmetry is broken, but the global color-rotation symmetry would not be broken under the vacuum expectation value (\ref{VEV}) since the non-gauged (right) index of $\phi $ may be rotated together.  Using the Faddeev-Popov method, the Lagrangian is
\bea
		{\cal L} &-& \frac{1}{2} \Big(\partial _{\mu } A^{a\mu } - i g M \tr\frac{\tau ^{a}}{2} (\phi  - \phi ^{\dagger }) \Big)^{2}		\nn
			&+& i \bar c^{a} \Big((\partial D)_{ab} + \frac{1}{2} g M \tr (\tau ^{a} \tau ^{b}\phi  + \tau ^{b}\tau ^{a} \phi ^{\dagger }) \Big) c^{b}.
\eea
It is invariant under the BRST transformation: \bea
		\delta _{B}A^{a}_{\mu } &=& \partial _{\mu } c^{a} + g f_{abc} A^{b}_{\mu } c^{c},	\nn
		\delta _{B} \phi _{ij} &=& i g c^{a}\tau ^{a}_{ik} \phi _{kj},		\nn
		\delta _{B} c^{a} &=& - g f_{abc} c^{b} c^{c}/2,		\\
		\delta _{B}\bar c^{a} &=& i B^{a},		\nn
		\delta _{B} B^{a} &=& 0.		\nonumber
\eea
This model gives the equation of motion (\ref{eqm}) as the classical counterpart.

Under the spontaneous symmetry breakdown of eq.(\ref{VEV}), the gauge bosons and ghosts acquire mass, and massive scalar particles appear due to Higgs mechanism.  We may take the mass of the Higgs particle large enough to make it physically irrelevant.

It is easy to show existence of mass gap in this model.  Here $ H + H_{m} $ is the full Hamiltonian with $\phi _{ij} \rightarrow  \phi _{ij} - v\delta _{ij}$ shifted variables, and $H_{m} = m^{2} A^{2}/2$.  For any eigenstate $|E\rangle $ with its energy $ E $, we have \be
		E = \langle E|H + H_{m}|E\rangle > \langle E|H_{m}|E\rangle ,
\ee
and the right-hand side should be a positive value unless the number of the gluon is zero. 
In general, \be
		|E\rangle = |n_{g} = 0\rangle + |n_{g} = 1\rangle + |n_{g} = 2\rangle  + \cdots ,
\ee
where $|n_{g}\rangle $ is a state with its gluon number $ n_{g} $.  Any energy eigenstate with quarks should include $n_{g} \ne 0$ states.  If it does not, \be
		(H_{0} + H_{\rm int} + H_{A})|E\rangle = E|E\rangle ,
\ee
where $ H_{\rm int}|E\rangle $ includes $ n_{g} = 1$ states and $ H_{A}|E\rangle $ should only include $n_{g} = 2$ and $n_{g} = 3$ states but the right-hand side can not include $n_{g} \ne 0$ states.

Due to nonbreaking of the BRST symmetry, the quarks are confined in this model because no colored state may appear as follows.  The color current is \be
		J_{\mu }^{a} =  f_{abc }A^{\nu b} F_{0\nu }^{c}(x) + j_{0}^{a} 
			+ (A_{\mu }\times B)^{a} - i (\bar c\times D_{\mu }c) + i (\partial _{\mu }\bar c\times c) ,
\ee
where $ j_{\mu }^{a}$ is the color current from the quarks and the Higgs field $ j_{\mu }^{a} = \bar q \gamma_{\mu } q + \phi (D_{\mu }\phi ^{\dagger }) + (D_{\mu }\phi )\phi ^{\dagger }$, where $ D_{\mu }\phi ^{\dagger } = \partial _{\mu }\phi ^{\dagger }+ i g \phi ^{\dagger }A_{\mu } $.  The Maxwell equation \be
		D^{\nu }F_{\nu \mu }^{a} + g j_{\mu }^{a} = \partial _{\mu } B^{a} - i g f_{abc} (\partial _{\mu }\bar c^{b}) c^{c} ,
\ee
may be written as $\partial ^{\nu }F_{\nu \mu }^{a} + g J_{\mu }^{a} = \{Q_{B}, D_{0}\bar c^{a}(x) \}$.
Under BRST formalism, physical states must be annihilated by the generator $Q_{B}$ \cite{KO78}.
Therefore $\langle g|\partial ^{\nu }F_{\nu \mu }^{a} + g J_{\mu }^{a}|f\rangle = 0$ for any physical states $|f\rangle $ and $|g\rangle $, which means that \be
		( \partial ^{\nu }F_{\nu \mu }^{a} + g J_{\mu }^{a} )|f\rangle = 0
\ee
inside the physical space.  The operator on the left-hand side is the generator of color gauge transformation, and then this equation means that only color-neutral state can appear as a physical state.

\section{Discussion}
In this letter, we presented a new classical solution of QCD and discussed possible relation to confinement.  Further, we found that confinement and mass gap may occur in a model with a explicit mass introduction with Higgs mechanism.

Dual Meissner effect, i.e. massiveness of magnetic field, has mainly been considered to be the mechanism that confines color charged particles, i.e.\ quarks.  However, the picture presented here shows that mass acquisition of electric field is rather appropriate to show the confinement string if the vector particle could acquire mass without breaking the conservation of color current.

This mechanism will be valid in the real QCD because 
the lattice QCD, which is equivalent to the real QCD, dynamically acquires mass from analytical and lattice studies\cite{BIMS, Nat, Agu}.
Further, we have another possibility that some of Higgsless massive vector field theories \cite{Su, Mof, Kon, TM} works. In that case, the analyses in the latter sections would be useful.


\end{document}